\documentclass[12pt]{iopart}

\usepackage{amssymb}
\usepackage{color}
\usepackage{graphicx}

\begin{document}

\title{Rotationally invariant family of L\'evy like random matrix ensembles}
\author{Jinmyung Choi and K.A. Muttalib}
\address{Department of Physics, University of Florida, Gainesville FL 32611-8440}
\ead{\mailto{jmchoi@phys.ufl.edu}, \mailto{muttalib@phys.ufl.edu}}
\begin{abstract}
We introduce a family of rotationally invariant random matrix ensembles characterized by a parameter $\lambda$. While $\lambda=1$ corresponds to well-known critical ensembles, we show that $\lambda \ne 1$ describes ``L\'evy like'' ensembles, characterized by power law eigenvalue densities.  For $\lambda > 1$ the density is bounded, as in Gaussian ensembles, but $\lambda <1$ describes ensembles characterized by densities with long tails. In particular, the model allows us to evaluate, in terms of a novel family of orthogonal polynomials, the eigenvalue correlations for L\'evy like ensembles. These correlations differ qualitatively from those in either the Gaussian or the critical ensembles.   
\end{abstract}


\maketitle

\section{Introdcution}
Gaussian Random Matrix Ensembles (RMEs) were proposed by Wigner about half a century ago to describe the statistical 
properties of the eigenvalues and eigenfunctions of complex many-body
quantum systems in which the Hamiltonians are  
considered only in a probabilistic way \cite{Wigner,Mehta}. Over the
past several decades, they proved to be a very useful tool in the studies of
equilibrium and transport properties of
disordered quantum systems, classically chaotic systems with a few degrees
of freedom, two-dimensional gravity,
conformal field theory and chiral phase transition in quantum
chromodynamics \cite{reviews}. This wide applicability results from certain universal
properties of the correlation of the eigenvalues known as the Wigner distributions, as opposed to the Poisson distributions that result from random sets of eigenvalues.

More recently, attempts have been made to construct generalized ensembles that show a crossover from a Wigner to a Poisson distribution as a function of a parameter, as seen in many physical systems \cite{crossover}. One such generalization is the family of `q-random matrix ensembles' (q-RMEs) \cite{qrme1,qrme2}. These  
were later shown to be models of `critical ensembles', 
with statistical properties different from the Gaussian RMEs and relevant for systems near a metal-insulator transition \cite{Kravtsov}. While there are other models that also describe critical statistics \cite{critical}, one advantage of the q-RMEs is that they are rotationally invariant, and therefore can be analytically studied in great detail by the powerful method of orthogonal polynomials \cite{Mehta}. In particular, the differences between the Gaussian and the critical RMEs can be traced to the differences in the asymptotic properties of the classical vs. q-orthogonal polynomials \cite{szego,ismail}, illustrating how the universality of the Gaussian RMEs breaks down and gives rise to a different kind of universality for the critical ensembles. 

In this work, we take the generalization one step further to include `L\'evy like' ensembles. L\'evy ensembles were introduced by Cizeau and Bouchaud (CB) \cite{CB}  where the matrix elements are drawn from a power
law distribution according to $P(H_{ij})\sim 1/|H{ij}
|^{1+\mu}, \;\; H_{ij} \gg 1, \; 0<\mu <1$. The eigenvalue density for such ensembles
falls off as $1/x^{1+\mu}$. Such matrices with a broad distribution of matrix elements 
have been studied in the context
of a wide variety of systems including financial markets, earthquakes, scale free networks, communication
systems etc. 
as well as quantum chaotic systems such as the Coulomb billiard and kicked rotor with singularities \cite{complex}.  The statistical properties of the
eigenvalues and eigenvectors of such matrices are, in general, quite different from the
universal properties of the Gaussian or the critical ensembles. Indeed, it has
been argued  \cite{CB} that matrices of the L\'evy type are relevant for describing
localization transition of interacting electrons in infinite dimensions. 
In particular, while the eigenvectors of the Gaussian RMEs are all extended,
the L\'evy type matrices contain signatures of mobility edges, separating
localized states from extended states within the spectrum.

However, the CB
model is very difficult to study because it is not
rotationally invariant. While eigenvalue densities for certain
L\'evy ensembles have been obtained analytically \cite{burda}, little
progress has been made in any systematic study of the eigenvalue \textit{
correlations} where novel universal features can be expected. 
In the present work we will consider `L\'evy like' ensembles, characterized by a power law eigenvalue density $1/x^{1+\mu}$, without specifying whether such a density arises from a distribution of the matrix elements as in the CB model. 
For this purpose, we introduce a rotationally invariant model characterized by a parameter $\lambda$. We show that $\lambda=1$ corresponds to a critical ensemble, while $\lambda > 1$ and $\lambda <1$ describe L\'evy like ensembles with bounded and unbounded (long tail) densities, respectively. In particular the model allows us to calculate the two-level kernel for L\'evy like ensembles, from which all correlations can be evaluated, in terms of a novel family of orthogonal polynomials. These correlations turn out to be qualitatively different from either the Gaussian or the critical ensembles.

Consider the set of all $N\times N$ Hermitian matrices $M$ (we will restrict ourselves to the unitary ensembles only) randomly chosen
with the following probability measure
$
P^V_N(M)dM\propto \exp[-Tr(V(M))]dM,
$
where $V(x)$ is a suitably increasing function of $x$, 
$Tr$ is the matrix trace and $dM$ the Haar 
measure. By going over to the eigenvalues-eigenvectors representation, it
can be shown that the joint probability distribution of the
eigenvalues $X=(x_i, i=1,2,\dots N)$ of the matrices can then be written
in the form \cite{Mehta}
\begin{equation}
P^V_N(X)\propto 
\prod_{1\leq i<j \leq N}^N (x_i - x_j)^2 \prod_{i=1}^N e^{-2V(x_i)}.
\end{equation}
Here the factor $\prod (x_i-x_j)^2$ arises from the Jacobian of  
a change of variables. 

The level correlations can be determined exactly by recognizing that the
distribution can be written as a product of
Vandermonde determinants of a set of (monic) polynomials $\phi_n(x)$ that 
are orthogonal with respect to the weight function $w(x)=e^{-V(x)}$ \cite{Mehta}, i.e.  
\begin{equation}
\int\limits_{-\infty}^{\infty} e^{-V(x)}\phi_n(x)\phi_m(x)dx
=\delta_{mn}.
\end{equation}
The main quantity of interest is the large N limit of the two-level kernel 
\begin{equation}
\label{eq16}
K^V_N(x,y)\equiv  e^{-(V(x)+V(y))/2}\sum_{n=0}^{N-1} \phi_n(x)\phi_n(y),
\end{equation}
from which all correlation functions can be obtained. The Gaussian 
RMEs follow when one chooses the `confinement potential' $V(x)=x^2$ that defines the Hermite polynomials. 
Since all `Freud-type' weight functions  $e^{-V(x)}$, with monotonically increasing \textit{polynomial} $V(x)$, lead to orthogonal polynomials with qualitatively similar asymptotic behavior in the large $N$ limit, they all share the same correlations as the Gaussian REMs. This is at the root of the wide applicability of the Gaussian RMEs. On the other hand, q-RMEs follow when one chooses $ V(x) \sim \ln^2 x$ for large $x$ \cite{qrme1}, which leads to the q-polynomials. All q-polynomials have similar large $N$ asymptotic behavior, which differ qualitatively from those of the Freud like classical polynomials. Thus all q-RMEs characterized by different q-polynomials share the same correlation properties as those of the critical ensembles, which are different from those of the Gaussian ensembles. 

\section{Model for L\'evy-like ensembles}
It turns out that all q-RMEs have density of eigenvalues falling off as $1/x$ \cite{mi}. Comparing this with the fact that  L\'evy-like ensembles have density falling off as $1/x^{1+\mu}$, this suggests that we consider a generalization of the q-RME model where the asymptotic $\ln^2 x$ behavior of the potential is extended to include other powers of the logarithm.
We therefore introduce a one parameter generailzation of the q-RMEs characterized by 
\begin{equation}
V (x) = \frac{1}{\ln(1/q)}[\sinh^{-1}x]^{1+\lambda}; \;\;\; \lambda > 0; \;\;\; q<1.
\end{equation}
For $\lambda$ = 1, this is a model for critical ensembles \cite{qrme2}, defined by the Ismail-Masson q-polynomials \cite{ismail}.  Since the asymptotic behavior of the potential is $V(x)\sim \ln^{1+\lambda}(x)$, any $\lambda \ne 1$ is qualitatively different from a q-ensemble.  However, there is no direct way to obtain orthogonal polynomials for any arbitrary non-trivial weight function. In particular, the orthogonal polynomials corresponding to $\lambda \ne 1$ are not known, and we can not write down the two-level kernel (Eq.~3) directly. 
We therefore follow an indirect but systematic method \cite{spm} that allows us to obtain the polynomials recursively for any given potential. It is well known that every orthogonal system of real valued polynomials satisfy a three term recursion relation \cite{szego, ismail}
\begin{equation}
x\phi_n(x)=\phi_{n+1}(x)+S_n\phi_n(x)+R_n\phi_{n-1}(x).
\end{equation}
Following Ref [15], we define a set of integrals 
\begin{equation}
Q_{n,m}\equiv\int_{- \infty}^{\infty}x^m e^{-2V(x)}\phi_n(x) dx,
\end{equation}
which in turn satisfy the recursion relation
\begin{equation}
Q_{n,m}=Q_{n-1,m-1}-S_{n-1}Q_{n-1,m}-R_{n-1}Q_{n-2,m}.
\end{equation}
Thus, the determination of the coefficients $R_n$ and $S_n$ necessary to calculate the polynomials of degree $n \le N-1$ requires only the knowledge of the $2N+1$ integrals  $Q_{0,m}$ for $m=0,1, \cdots 2N$.  
\begin{figure}[tbp]
\begin{center}
\includegraphics[angle=0, width=0.35\textheight]{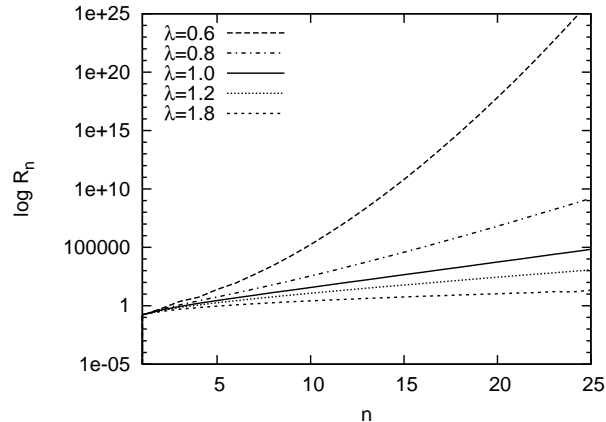}
\end{center}
\caption{Log $R_n$ as a function of $n$ for different values of $\lambda$. Solid line corresponds to the critical ensemble $\lambda=1$, which separates two qualitatively distinct classes $\lambda >1$ and $\lambda <1$.} \label{Rn}
\end{figure}

\section{Results}
\begin{figure}[tbp]
\begin{center}

\includegraphics[angle=0, width=0.35\textheight]{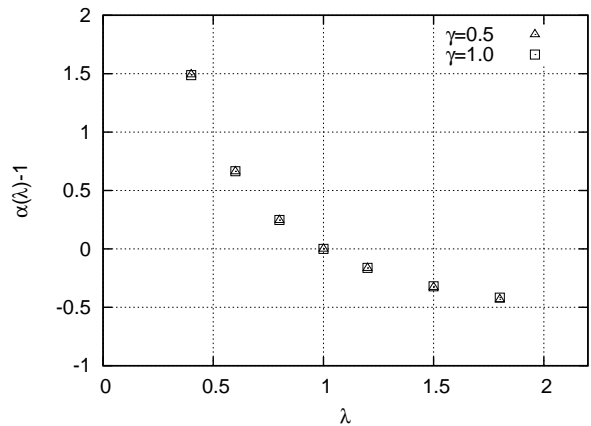}
\end{center}
\caption{The exponent $\alpha(\lambda)$ as a function of $\lambda$ for two different values of $\gamma\equiv\ln(1/q)$. $\lambda=1$ corresponds to the q-polynomials describing the critical ensembles.} \label{alpha}
\end{figure}

We first note that for the present model, because we choose the potential to be symmetric, all $S_n=0$. Therefore, $R_n$ completely determines the polynomials. For simplicity, we will use a fixed value of the parameter $\ln(1/q)\equiv \gamma=0.5$ except in Figure~\ref{alpha} where we also show results for $\gamma=1$. Figure~\ref{Rn} shows the $n$-dependence of $R_n$ for different values of $\lambda$ obtained by evaluating the $Q_{0,m}$ numerically. For all $\lambda$, the large $n$ behavior of $R_n$ has the form 
\begin{equation}
R_n\propto q^{-n^{\alpha(\lambda)}}.
\end{equation}
In contrast, for all Freud like classical orthogonal polynomials, $R_n\propto n$. Figure~\ref{alpha} shows the $\lambda$ dependence of the exponent $\alpha$. 

For $\lambda=1$, $\alpha(\lambda) =1$, which recovers the known recursion relation\footnote{Exact expression for all $n$ is $R_n=(q^{- n}-1)/4$, see \cite{ismail}.} for the q-orthogonal polynomials. On the other hand for $\lambda >1$, $\alpha(\lambda) <1$, while for $\lambda <1$, $\alpha(\lambda) >1$. Note that the distribution of the zeros of the polynomials are determined by the exponent $\alpha$, such that when $\alpha =1$, the logarithm of the zeros of the corresponding Ismail-Masson q-polynomials are uniformly distributed. For $\alpha <1$ the zeros are `bunched together' while for $\alpha >1$ they are `unbunched', highlighting their differences from the q-polynomials. 
Thus, $\lambda \ne 1$ defines a novel family of orthogonal polynomials that generalizes the q-polynomials. However, the asymptotic properties of these `generalized q-polynomials' are not known.

\begin{figure}[tbp]
\begin{center}
\includegraphics[angle=0, width=0.35\textheight]{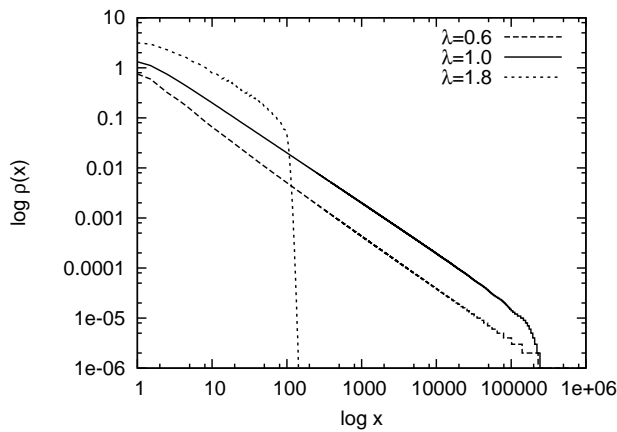}
\end{center}
\caption{Density of eigenvalues for different values of $\lambda$. $\lambda=1$ is the critical ensemble. For $\lambda >1$, the density has a sharp edge, while for $\lambda<1$ there is a true `Levy-like' long tail. } \label{density}
\end{figure}

The density of eigenvalues $\rho(x)=K_N^V(x,x)$ can now be obtained for different values of $\lambda$ from Eq.~3 by summing the products numerically. The results are shown in Figure~\ref{density}.  
As already known, the density for $\lambda=1$ falls off as $1/x$. 
For $\lambda \ne 1$, the density falls off as $1/x^{1+\mu(\lambda)}$, as expected for L\'evy-like ensembles. However, for $\lambda >1$, $\mu(\lambda) <0$ and the density falls off slower than $1/x$; the normalization condition (that there are $N$ eigenvalues in total for a matrix of size $N$) then forces the density to have a sharp edge, similar to the edge of the semicircular density of the Gaussian RMEs. Thus, even though all $\lambda \ne 1$ have power law densities, the long tail characteristic of a L\'evy ensemble is cut off for $\lambda > 1$ by a sharp edge. For $\lambda <1$, $\mu(\lambda) > 0$ and the density has true power law tails.
It should be noted that nonextensive ensembles \cite{nonextensive} characterized by one parameter $q$ also show a similar power-law behavior; for $q>1$, the distributions of eigenvalue density show true long tails and for $q<1$, the distributions have compact support. However, the parameter $q$ in this case depends on the dimensionality N of the ensemble such that in the large $N$ limit where universal behavior is expected, the maximum $q$ allowed for the nonextensive ensembles approaches unity.

\begin{figure}[tbp]
\begin{center}
\includegraphics[angle=0, width=0.35\textheight]{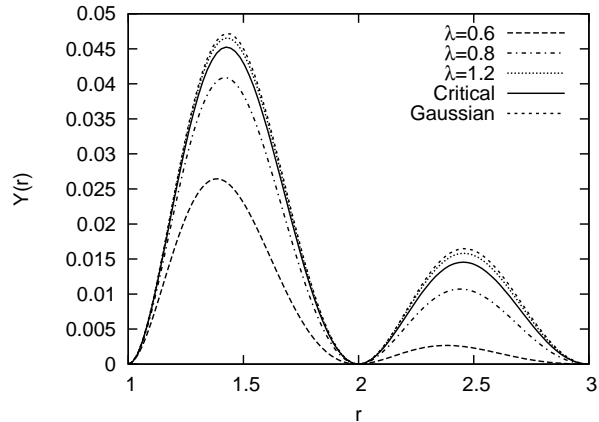}
\end{center}
\caption{The cluster function $Y(r)$ for L\'evy-like ensembles for different values of $\lambda$. The dotted and the solid lines agree with the known results for the Gaussian and the critical ensemble (with $\gamma =0.5$), respectively.} \label{cluster}
\end{figure}

Our results now allow us to obtain exact correlations for L\'evy like ensembles. Using Eq.~3 for the two-level kernel we calculate the `unfolded' cluster function\footnote{We restrict ourselves to a range of parameters where the kernel can be considered translationally invariant.} $Y(r)\equiv |K^V_N(r)|^2$ where the variable $r$ is such that the density is uniform and unity: $\bar{\rho}(r) =1$. It is only in this rescaled variable $r$ that the universality of a given ensemble is revealed if it exists. The results are shown in 
Figure~\ref{cluster}. Note that cluster functions for $\lambda >1$ are intermediate between the Gaussian ($Y^G$) and the critical ($Y^C$) ensmbles  
\begin{equation}
\label{Y2}
Y^{G}=\left[\frac{\sin(\pi r)}{\pi r}\right]^2; \;\;\; Y^{C}=\left[\frac{\gamma}{2\pi}\frac{\sin(\pi r)}{\sinh [(\gamma/2)r]}\right]^2; 
\end{equation}
but for $\lambda <1$ it is qualitatively different, with peak positions of the lobes shifting towards smaller $r$ values with decreasing $\lambda$. This suggests that $Y$ consists of a more general $\lambda$-dependent argument as well as power of the $\sinh$ function that reduces to $Y^C$ in the limit $\lambda=1$. Note that our results are limited to finite system size N only, which makes any attempt to obtain the true asymptotic behavior of $Y$ unreliable. Nevertheless, with sufficiently large N, we can clearly demonstrate that the correlation functions obtained from $Y$ for $\lambda \neq 1$ are qualitatively different from both the Gaussian and the critical ensembles. 

\begin{figure}[tbp]
\begin{center}
\includegraphics[angle=0, width=0.35\textheight]{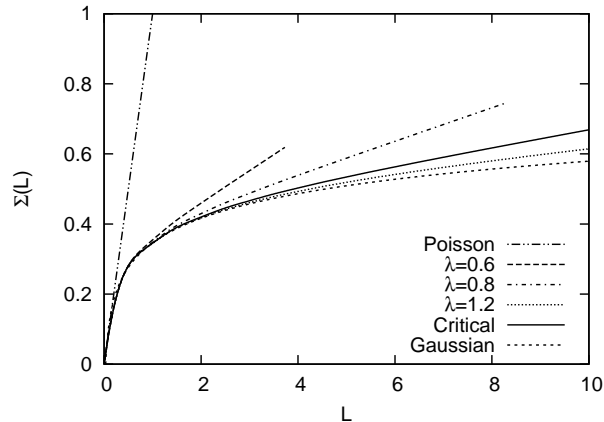}
\end{center}
\caption{Number variance $\Sigma(L)$ for L\'evy like ensembles for different values of $\lambda$} \label{variance}
\end{figure}

As an example of the correlation functions obtained from the cluster function, we evaluated the number variance $\Sigma(L)$ within a range $L$ shown in Figure~\ref{variance}. For $\lambda >>1$, the number variance tends to that of the Gaussian RMEs, as expected, due to the absence of long tails. On the other hand for $\lambda <1$, the number variance shifts towards the uncorrelated Poisson distribution. For $\lambda <1$, $\Sigma(L)$ for large $L$ seems linear in $L$, as it is for the critical as well as Poisson cases, although the slope depends on $\lambda$ in a nontrivial manner. We have also evaluated the gap function which, for critical ensembles, falls off slower compared to the Gaussian RMEs. We find that for $\lambda >1$ the gap function is in between the critical and the Gaussian RMEs, while for $\lambda <1$ it falls off slower than the critical ensemble. Again, finite system size prevents us from obtaining a reliable $\lambda$ dependence of the asymptotic behavior. Determination of the true asymptotic behavior of these correlation functions will require knowledge of the asymptotic properties of the novel orthogonal polynomials introduced here. 

\section{Summary \& Conclusion}
In summary, we have introduced a rotationally invariant model of random matrix ensembles by defining a confinement potential $V(x)$ that includes a parameter $\lambda$. We obtained the orthogonal polynomials associated with the weight function $w(x)=e^{-V(x)}$ for different values of $\lambda$ by using a recursive method. For $\lambda=1$, we recover the well known q-polynomials that describe the critical ensemble, with eigenvalue density $\rho(x)\propto 1/x$. For $\lambda \ne 1$, we obtain a novel family  of `generalized q-polynomials'. While $\rho(x)\propto 1/x^{1+\mu(\lambda)}$ for all $\lambda \ne 1$, only $\lambda <1$ correspond to L\'evy like ensembles with true long tails. 
We use the generalized q-polynomials (for finite N) to evaluate the two-level kernel from which all eigenvalue correlations can be obtained. We find that for $\lambda \ne 1$, the correlations are of  novel types, differing from both the Gaussian and the critical ensembles. Clearly, studies of the asymptotic properties of the generalized q-polynomials introduced above would be very useful in better understanding the properties of L\'evy like ensembles. 

Although we have considered only the variation with respect to the parameter $\lambda$, the model has another parameter, $q$, which we kept fixed. We have confirmed that the two parameters are independent, so that a single effective parameter can not describe the separate dependence of the two parameters. It would be important to find out if the q-parameter can describe e.g. any ``transition'' within the L\'evy like ensembles, as argued in Ref. \cite{CB}. While
the mobility edge or the localization transition has clear meaning in the
context of electron transport in disordered systems, it would be very interesting if it exists in L\'evy like ensembles, in the context of complex systems like
earthquakes and scale-free networks.

\end{document}